\documentclass{PoS}

\title{Solving the polarization problem in ALMA-VLBI observations}

\ShortTitle{Polarization in ALMA-VLBI}

\author{\speaker{Ivan Mart\'i-Vidal}\thanks{The author thanks the COST Action MP1104 for travel support to the MPIfR}, John Conway, Michael Lindqvist\\
        Dpt. of Earth and Space Sciences, Chalmers University of Technology\\
        Onsala Space Observatory, SE-43992, Onsala (Sweden)
        E-mail: \email{mivan@chalmers.se}}

\author{Alan Roy, Walter Alef, J. Anton Zensus\\
        Max-Planck-Institut f\"ur Radioastronomie\\
        Auf dem H\"ugel 69, DE-53121 Bonn (Germany)}

\abstract{The Atacama Large mm-submm Array (ALMA) is, by far, the most sensitive mm/submm telescope in the World. The ALMA Phasing Project (APP) will allow us to phase-up all the ALMA antennas and use them as one single VLBI station. This will be a key component of the Event Horizon Telescope (EHT), a Global VLBI array at millimeter wavelengths. A problem in the
APP is the calibration and conversion of the polarization channels. Most VLBI stations record their signals in a circular basis, but the ALMA receivers record in a linear basis. The strategy that will be followed in the phased-ALMA VLBI observations will be to correlate in a ``mixed'' basis (i.e., linear versus circular) and convert the visibilities to a pure circular basis after the correlation. We have developed an algorithm to perform such a polarization conversion of the VLBI visibilities. In these proceedings, we present the basics of this algorithm and discuss on the polarization conversion in the general case where single dishes (besides phased arrays) record with linear receivers in VLBI observations. We show some results of our algorithm applied to realistic simulations, as well as a test with real VLBI observations at 86\,GHz between the Onsala radiotelescope (recording in linear basis) and the Effelsberg radiotelescope (recording in circular basis).}

\FullConference{12th European VLBI Network Symposium and Users Meeting,\\
		7-10 October 2014\\
		Cagliari, Italy}

\begin{document}

\section{Introduction}

Most heterodyne receivers register the signals in a linear polarization basis. The receiver's frontend can be understood as two orthogonal dipoles, one horizontal (the X axis) and one vertical (the Y axis), in the frame of the antenna mount. However, Very-Long-Baseline-Interferometry (VLBI) observations involve baselines so long, that the parallactic angles, $\psi$, of two antennas pointing to the same source can be quite different (assuming that the mounts of the antennas are alt-azimuthal). This orientation issue in VLBI can originate many problems if the signals are recorded in a linear polarization basis, since the $\psi$ correction will mix, in that case, the X and Y signals {\em in different ways} for the different antennas. Hence, both channels X and Y have to be detected and recorded with similar system temperatures, $T_{sys}$, and the amplitude calibration must be well known before aapplying any $\psi$ (and/or phase) correction.

However, if the signals are recorded in a circular polarization basis (i.e. dividing the light into right-hand, RCP, and left-hand, LCP, circular polarizations), all these problems are solved. Circular receivers can be built by just adding a quarter waveplate (or an equivalent device) at the frontend of the receiver, so that the incoming circular polarization is converted into linear, and then detected by the dipoles. In this R/L basis, the $\psi$ correction becomes a mere phase addition to RCP and LCP, so there is no need of having neither similar $T_{sys}$ nor known amplitude correction factors in the initial data calibration. In addition to this, if any of the channels, RCP or LCP, fails to be recorded (or if it is of bad quality), it is still possible to use the other channel for the science analysis. This is not possible if the recording is performed in the X/Y basis, since both channels are always needed for the $\psi$ correction (especially if $\psi$ is not the same for all antennas).

Circular-polarization receivers have, though, some disadvantages. The purity of the polarization is usually lower (there is more polarization leakage between the channels) and the effective bandwidth is narrower. Quarter waveplates are designed to be optimum at a given frequency, but their performance degrades as the signal frequency departs from that used in the waveplate design. These disadvantages make linear-polarization receivers the optimum choice for interferometers with wide-band receivers and high polarization purity, like ALMA. 

Hence, ALMA uses receivers that record the signal in a linear, X/Y, basis, whereas VLBI stations mostly record the signals in a circular, R/L, basis. As a consequence, when the phased ALMA will be used as a VLBI station, the visibilities that will be correlated against the other VLBI stations will be computed in a ``mixed'' polarization basis (i.e., linear for ALMA, circular for the other stations). These visibilities have to be converted into a pure circular basis, before any analysis or calibration can be performed. In these proceedings, we briefly describe our approach to calibrate and convert the ALMA-VLBI ``mixed-polarization'' visibilities into a pure circular basis.

\section{Visibility matrix in a mixed-polarization basis involving a phased array}

Let $v^a_k$ be the voltage registered by an antenna $a$ at time $t_k$ in any given polarization basis. Antenna $a$ may be, in our case, a stand-alone VLBI station with a circular-polarization feed. Let $v^i_k$ be the voltage at time $t_k$ registered by the $i$-th antenna of a phased array (which will be named $b$) in another polarization basis. Since the phased signal of $b$ at time $t_k$ is the addition of the voltages $v^i_k$ of all the phased antennas, the corresponding element of the Visibility matrix (see Smirnov 2011a) between the stations $a$ and $b$ will be

\begin{equation}
\left< v^a v^b\right> = \left< v^a \sum_i{v^i}\right> = \sum_i{\left<v^a v^i \right>}. 
\label{Lab1}
\end{equation}

\noindent That is, given that the correlation operator is linear, the visibility is equal to the sum of the correlations between antenna $a$ and each individual antenna of the phased array. Taking into account all the polarization products, we can build the full Visibility matrix in the same way, i.e.

\begin{equation}
V(a,b) = \sum_i{V(a,i)}.
\label{Lab2}
\end{equation}

In the absence of noise, and if the signals are perfectly calibrated, all the matrices $V(a,i)$ should be numerically equal (as long as the synthesized VLBI field-of-view is much smaller than the synthesized resolution of the phased array). 



\section{The effect of different antenna gains in the phased elements}

The different antennas of the phased array are affected by slightly different gains, bandpass responses, and polarization leakage (D-term factors). The observed Visibility matrix, $V^{obs}(a,b)$, will then be related to the perfectly-calibrated Visibility matrix, $V(a,b)$, by the equation

\begin{equation}
V^{obs}(a,b) = V(a,b)\,K,
\label{calEq1}
\end{equation}

\noindent where the calibration Jones matrix, $K$, can be a function of frequency and time, and is related to all the calibration matrices (i.e., gain, bandpass, D-terms, etc.) of all the phased-array antennas. Using the algebra linearity, we can also write

\begin{equation}
V^{obs}(a,b) = \sum_i{V(a,i)\,K_i},
\label{calEq2}
\end{equation}

\noindent where antenna $i$ is corrupted by an unknown overall gain, $K_i$. 
Given that the uncorrupted visibility $V(a,i)$ is assumed to be noise-free, perfectly calibrated, and independent of $i$, we can put $V(a,i)$ out of the sum and write (see Eq. \ref{calEq1})

\begin{equation}
K = \left<K_i\right>,
\label{KavgEq}
\end{equation}

\noindent where $\left<...\right>$ is the averaging over the antennas of the phased array. Our objective is to find out the $K$ matrix of Eq. \ref{calEq1}, using all the observables available. 
During the VLBI observations, the ALMA correlator will be producing all the cross-correlations among ALMA antennas. These products will allow us to find the gains that calibrate each individual ALMA antenna (just from a standard ALMA data reduction). Hence, it will be possible to compute the $K$ matrix exactly and fully calibrate the mixed-polarization Visibility matrix (Eq. \ref{calEq1}) before the conversion of the data into a pure circular basis. 

Let $B^i_x$ and $B^i_y$ be the bandpass gains of the $i$-th antenna in the phased array for polarization X and Y, respectively. Let also $\alpha_i = \exp{(j\Delta\phi_i)}$ be the relative phase-gain between X and Y, for the same antenna, and $D^i_x$ and $D^i_y$ be the D-terms. 
Applying the corresponding Jones matrices (Smirnov 2011b), the exact $K$ matrix for the full phased array becomes

\begin{equation}
K = \left( \begin{array}{cc}
\left<B_x\right> & \left<D_x\,B_x\right> \\
\left<D_y\,B_y\,\alpha\right> & \left<B_y\alpha\right> 
\end{array} \right),
\label{Lab4}
\end{equation}

\noindent where $\left<...\right>$ represents the averaging over all the antennas. The calibration matrix to be multiplied by the observed visibility matrix will be the inverse of $K$ (see Eq. \ref{calEq1}). Time-dependent amplitude and phase gains can also be trivially added to Eq. \ref{Lab4} (just like the bandpass gains). Once the visibilities are calibrated with this matrix, the conversion from the ``mixed'' basis into pure circular basis is trivial. It is only necessary to multiply the visibility matrix by the ``hybrid'' linear-to-circular polarization matrix, which we call $C$:

\begin{equation}
C = \left( \begin{array}{cc}
1 & i \\
1 & -i 
\end{array} \right).
\label{Lab5}
\end{equation}

\section{Special case: single dish with a linear polarization receiver}

If the station with linear receiver is a single dish, we have no way to compute the $K$ matrix from any linear-linear cross-correlation. Although the conversion with the $C$ hybrid matrix is still possible, not correcting for the X and Y gains beforehand may lead to strong time-dependent leakage-like effects in the circular-circular visibilities after the conversion. Fortunately, we can still use the information encoded in the ``mixed'' visibilities to perform a proper calibration. If $V_{rr}$ and $V_{ll}$ are visibilities for the parallel-hand correlations RR and LL, respectively, then $V_{rr}/V_{ll} = 1$, regardless of the source structure (and provided that the source is not circularly polarized). Let us now suppose that the first antenna of the baseline observes with a linear receiver. Each polarization channel, X and Y, is affected by a different gain, $G_x$ and $G_y$. However, {\em if the polarization leakage is negligible}, the only quantity that is important for the calibration before the polarization conversion is the ratio of gains (i.e., the cross-polarization gain), $G_x/G_y = G_{x/y}$, which can be fortunately derived in an easy way. If we write $V_{rr}/V_{ll}$ in terms of the mixed-polarization visibilities and the gain ratios, we can define a norm, $\chi^2$, whose minimum will determine the gain ratios:

\begin{equation}
\chi^2 = \sum_k{\omega_k\left[ \frac{V_{xr}G^{-1}_{x/y} - i\,V_{yr}}{V_{xl}G^{-1}_{x/y} + i\,V_{yl}}(G^*_{L/R})^{-1} - 1\right]^2},
\end{equation}

\noindent where $\omega_k$ is the weight of the $k$-th visibility matrix and $G_{L/R}$ is the ratio of gains of the antenna(s) with circular receivers. An advantage of this equation is that $G_{x/y}$ shall be stable over time, so we can apply long integration times to derive $\chi^2$ and use this approach even with weak sources. If the leakage is not negligible, more complex expressions for $\chi^2$ are obtained.

\section{Implementation of the polarization conversion}

We have developed {\em PolConvert}, the calibration/conversion software that will be used in mm-VLBI observations with the phased-ALMA. The program reads the VLBI data in standard FITS-IDI format 
and converts the visibilities to a pure circular basis. 
It can optionally read CASA calibration tables to compute the $K$ matrix(ces) (Eq. \ref{Lab4}), and correct the VLBI visibilities (Eq. \ref{calEq1}) before the polarization conversion. 
The CASA tables currently supported in PolConvert are: G Jones (i.e. gain and X-Y phase), B Jones (i.e., bandpass), K Jones (i.e., delay)\footnote{Not to be confused with our full calibration $K$ matrix}, and D Jones (i.e., polarization leakage). As noted above, all these tables are optional for the calibration/conversion.

\subsection{Testing the algorithm. Simulations}

We have developed a simulator program to test the performance of the calibration/conversion algorithm implemented in PolConvert. Our 
program 
generates synthetic data under realistic conditions. It accounts for noise from the receivers, signal quantization (amplitude and time), and corruption effects (atmosphere). The simulator creates a synthetic set of ALMA-only cross-correlations (in a pure linear basis) and the corresponding phased-up data streams, which are cross-correlated with the streams simulated for a VLBI station recording in a circular basis. We then calibrate the ALMA-only visibilities following the usual ALMA reduction procedures, to derive the gain matrices of all the ALMA antennas. Finally, we let {\em PolConvert} compute the $K$ matrix, apply it to the VLBI fringes, and rewrite them in a pure circular basis.

\begin{figure*}[ht!]
\centering
\includegraphics[width=15cm]{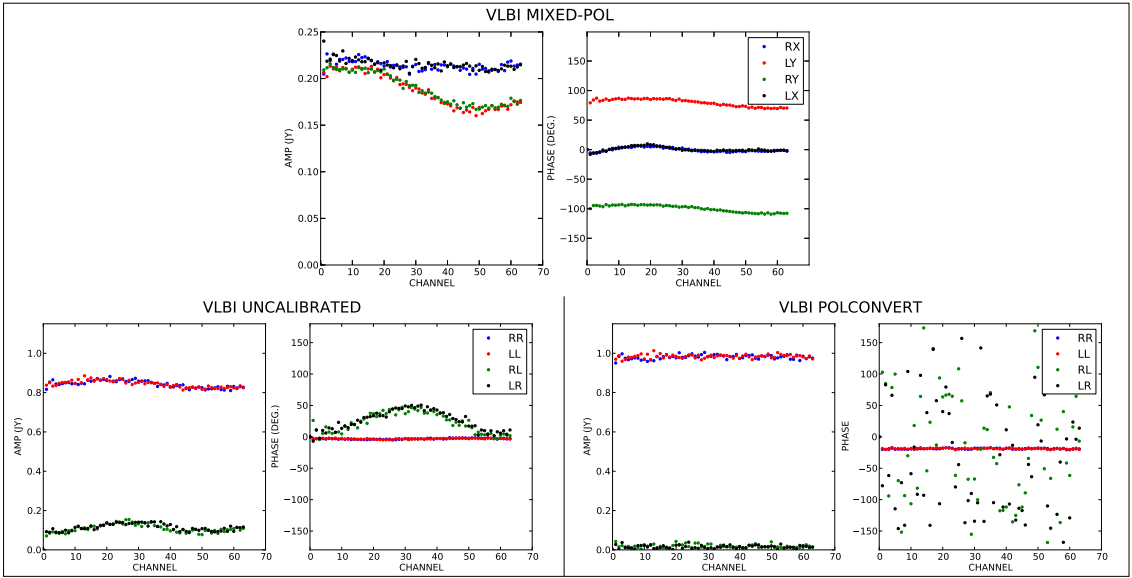}
\caption{Simulated ALMA-VLBI visibility of an unpolarized source in the ``mixed'' basis (top) and in pure circular basis (bottom). Bottom left, with no $K$-matrix correction. Bottom right, after a full calibration and conversion with {\em PolConvert}.}
\label{VLBIFig}
\end{figure*}

In Fig. \ref{VLBIFig}, we show an example of one of our simulations. In this case, the source is point-like and unpolarized, with a flux density of 1\,Jy. The visibilities in the ``mixed'' basis are shown at the top of the figure. Strong bandpass effects and phase offsets can be clearly seen. At bottom left, we show the visibilities converted to circular basis, but without pre-applying any correction (i.e. $K$) matrix. This figure illustrates the data quality that we would obtain if the polarization conversion were applied either at the recording or at the correlation stage (i.e., {\em before} knowing the true gains of each individual ALMA antenna). There is a large polarization leakage left in the cross-hand correlations (about 10\% of the source flux density!). Finally, at bottom right we show the visibilities resulting from the full calibration and conversion using {\em PolConvert}. All bandpass and leakage artifacts disappear and the resulting visibility is fully corrected.

\subsection{Testing the algorithm. Real VLBI observations}

On 22 May 2014, we performed a fringe-test VLBI session at 86\,GHz, using the Onsala 20\,m telescope and the Effelsberg radiotelescope. 
During part of this session, while observing the source OJ\,287, the staff at Onsala removed the quarter waveplate at the receiver's frontend, hence sending the sky radiation in linear basis directly into the receiving dipoles. Effelsberg continued data recording with no changes in its receiver. Hence, during that time window, Onsala was recording the signal in linear (X/Y) basis, while Effelsberg was recording in circular (R/L) basis. We correlated these data with DifX, which resulted in fringes in a ``mixed'' polarization basis, 
and performed a run of PolConvert, to obtain the fringes in a pure circular basis.

\begin{figure*}[ht!]
\centering
\includegraphics[width=16cm]{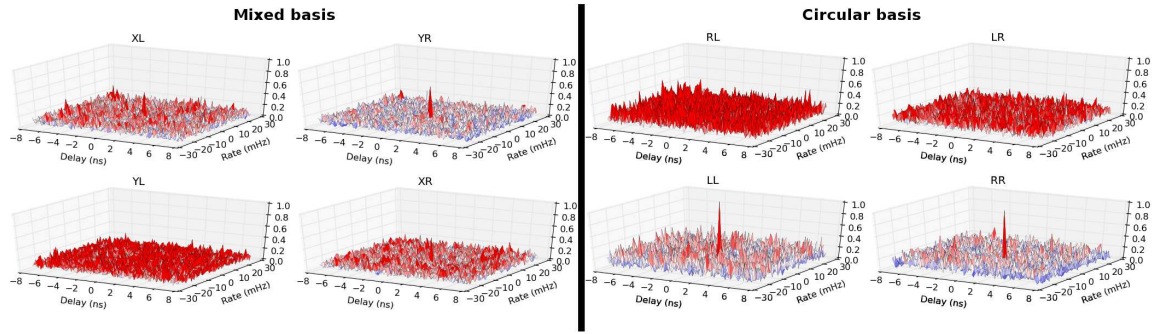}
\caption{VLBI visibility at 86\,GHz between Onsala and Effelsberg. Right, in ``mixed'' basis (Onsala recording in linear basis, Effelsberg in circular). Left, after running {\em PolConvert}.}
\label{VLBIFig2}
\end{figure*}

In Fig. \ref{VLBIFig2}, we show the amplitude fringes of this observation in delay-rate space. At left, the fringes are shown in the ``mixed'' basis; at right, in pure circular basis. Only weak detections are found in all four correlation products for the ``mixed'' basis. However, the signals are very clear in the parallel-hand correlations of the circular basis (RR and LL). The cross-hand correlations (RL and LR) do not show any clear detection, as it should be the case for an unpolarized source.

\section{Conclusions}


We have developed {\em PolConvert}, the software for the calibration/conversion of the ALMA mm-VLBI visibilities. We have successfully tested {\em PolConvert} with realistic synthetic data and with real state-of-the-art VLBI observations, performed in mixed polarization basis. We have shown that it is possible to fully calibrate and convert the VLBI visibilities related to ALMA. If the station with linear receivers is a single dish, we propose a special calibration algorithm, based on visibility ratios, to properly convert the visibilities to a circular basis.





\end{document}